\documentstyle[10pt]{article}
\normalbaselines
\begin{document}

\baselineskip=12pt

\bibliographystyle{unsrt}
\vbox{\vspace{6mm}}

\begin{center}
{\large\bf Parametric image amplification in optical cavities}
\end{center}

\bigskip 

\begin{center}
{\bf S. Mancini\footnote{Presenting author.},
A. Gatti}
\end{center}

\begin{center}
INFM-Dip. di Fisica, 
Universit\`a di Milano,
Via Celoria 16, 20133 Milano, Italy
\end{center}

\begin{center}
{\bf L. A. Lugiato}
\end{center}

\begin{center}
INFM-Dip. di Scienze, Universit\`a
dell'Insubria, Via Lucini 3, 22100 Como, Italy
\end{center}

\bigskip

\begin{abstract}
We show the possibility 
of noiseless amplification of an optical image in cavities 
containing a 
parametric oscillator. We consider a confocal
ring cavity with plane mirrors and compare
with the case of planar cavity.
In the latter case the system 
operates with severe spatial limitations, while in the  
confocal case, 
there is the possibility of preserving the 
signal-to-noise ratio while amplifying uniformely the entire 
image. 
\end{abstract}

\section{Introduction}

It is well known that phase-insensitive amplifiers introduce at least 3 dB extra noise in the
output, whereas phase-sensitive amplifiers may preserve the input signal-to-noise ratio, and in
this sense are ``noiseless" \cite{noiseless}.
A spectral analysis of some cavity-based noiseless amplifiers has been given in Ref.\cite{PLF94},
but these investigations are carried out exclusively in the time domain, while spatial aspects are
neglected by introducing the plane wave approximation,
i.e. by considering only one spatial
mode. 

However, the spatial domain is relevant for the subject of noiseless
amplification. Indeed, many areas of physics would benefit from having a possibility of noiseless
amplification of faint optical images. 
Here, based on this motivation, 
we analyse the parametric image amplification in two different cavity geometries
\cite{KL95,our}.

\section{The Optical Scheme}

A possible realization of a parametric image amplifier is shown in Fig.1.
A faint image which is to be amplified is located in the object plane $O$.
This image is projected by a lenses system in the input plane of a ring-cavity 
degenerate optical parametric amplifier.
The amplified image from the output plane is then projected in the image plane $I$ 
by another lenses system. Each  lens has a focal length $f$.
The presence of the dashed part in the scheme depends on the
geometry one wants to consider, as we shall see.
Instead, the presence of a pupil ${\cal P}$ of finite area $S_p$, is necessary for 
evaluation of the noise properties.
Let us now intruduce the two-component transverse wave vector
${\vec q}\equiv (k_x,k_y)$ and position vector ${\vec\rho}\equiv (x,y)$.
Moreover, let $a({\vec\rho}\,,t)$, $a^{\dag}({\vec\rho}\,,t)$,
and  $e({\vec\rho}\,,t)$, $e^{\dag}({\vec\rho}\,,t)$ the 
photon annihiliation and creation operators of the field in the
object plane and image plane respectively.
We shall assume that the field in the object plane is in a coherent state with 
real amplitude $s({\vec\rho}\,)$ modulated (and even) in space.
The observable, in the image plane, is the surface photocurrent density 
$i({\vec\rho}\,,t)=\eta\langle e^{\dag}({\vec\rho}\,,t)
e({\vec\rho}\,,t)\rangle$, with $\eta$ the photodetection efficiency. 
However, the quantity of interest for us is the number of 
photodetections $N_I({\vec\rho}\,,t)$ registered  by the pixel,
of area $S_d$, centered at 
the point $\vec\rho$ in the image plane in the time window 
$[t-T_d/2,t+T_d/2]$, i.e.
\begin{equation}\label{NI}
N_I({\vec\rho}\,,t)=\int_{S_d}\, d{\vec\rho}\,'\,
\int_{T_d}\, dt'\, i({\vec\rho}\,'\,,t'\,)\,.
\end{equation}
We shall consider the mean number $\langle N_I({\vec\rho}\,,t)
\rangle$ of registered electrons as the amplified signal of 
our scheme. Its variance characterizes the noise properties 
of the 
image \begin{equation}\label{DeltaNI}
\langle\Delta N_I^2({\vec\rho}\,,t\,)\rangle=
\int_{S_d}\, d{\vec\rho}\,'\,
\int_{T_d}\, dt'\, 
\int_{S_d}\, d{\vec\rho}\,''\,
\int_{T_d}\, dt''\,
\langle\mbox{$\frac{1}{2}$}\{\delta i({\vec\rho\,}',t'),
\delta i({\vec\rho\,}''\,,t'')\}_+\rangle\,.
\end{equation} 
The power signal-to-noise ratio (SNR) of the amplified image  
is given by \begin{equation}\label{RI}
R_I=\langle N_I({\vec\rho}\,,t\,)\rangle^2/
\langle\Delta N_I^2({\vec\rho}\,,t\,)\rangle\,.
\end{equation} 
Equivalent quantities, 
$\langle N_O({\vec\rho}\,,t\,)\rangle$,
$\langle\Delta N_O^2({\vec\rho}\,,t\,)\rangle$
and $R_O$, can be built up in the object plane.  
Finally, by definition, the noise figure is
\begin{equation}\label{F}
F=R_O/R_I\,,
\end{equation}
and we refer to the situation of $F\to 1$ as the noiseless amplification.
In order to investigate this possibility we have to
express the image field in terms of object field.

\section{Plane Cavity}

We first consider the case of planar cavity \cite{KL95}. Hence, 
the dashed part in Fig.1 has to be neglected. 
The behavior of the slowly varying part of the field 
operator $b({\vec\xi}\,,t)$ inside the cavity is described by the following
Langevin equation
\begin{equation}\label{Langb1}
\partial_t b({\vec\xi}\,,t)-ic\nabla_{\perp}^2/2k=
-\gamma\left[(1+i\Delta)
b({\vec\xi}\,,t)-{\cal A}_p\, 
b^{\dag}({\vec\xi}\,,t)\right]+\sqrt{2\gamma}\,b_{in}({\vec\xi}\,,t)\,,
\end{equation}
where $k=2\pi/\lambda$ is the wave number, $\gamma$ is the cavity decay rate, and 
$\Delta$ is the cavity detuning.
${\cal A}_p$ is the constant of parametric interaction proportional to the amplitude 
of the pump field taken as a classical quantity.
The symbol $\nabla_{\perp}^{2}$ denotes the transverse laplacian.
The above equation can be solved by means of spatio-temporal 
Fourier transformations;
furthermore, with the aid of transformations at lenses
\begin{equation}\label{lenses1}
b_{in}({\vec\xi}\,,t)=
\int
\frac{d{\vec\rho}}{\lambda f}\,
a({\vec \rho}\,,\Omega)
e^{
-i\frac{2\pi}{\lambda f}{\vec\rho}\,\cdot{\vec\xi}}\,,
\quad
e({\vec\rho}\,,t)=\int
\frac{d{\vec\xi}}{\lambda f}\, {\cal P}({\vec\xi}\,)\,
b_{out}({\vec\xi}\,,t)
e^{
-i\frac{2\pi}{\lambda f}{\vec\rho}\,\cdot{\vec\xi}}
\end{equation}
we arrive at
\begin{equation}\label{eofa1}
e({\vec\rho}\,,\Omega)=\frac{1}{\lambda f}\int
d{\vec\rho}\,'\, \wp({\vec\rho}\,-{\vec\rho}\,'\,)\,
\left[U({\vec\rho}\,'\,,\Omega)a({\vec\rho}\,'\,,\Omega)
+V({\vec\rho}\,'\,,\Omega)a^{\dag}({\vec\rho}\,'\,,-\Omega)
\right]
\end{equation}
where $\wp$ is the impulse response function, 
i.e. the Fourier transform 
of the pupil function ${\cal P}$
(for an infinitely large pupil $\wp({\vec\rho}\,)=\lambda 
f\delta({\vec\rho}\,)$).
The coefficients $U,V$ are 
\begin{eqnarray}
U({\vec\rho}\,,\Omega)&=&\frac{
[1-i\delta({\vec\rho}\,,\Omega)]
[1-i\delta({\vec\rho}\,,-\Omega)]+{\cal A}_p^2}
{[1+i\delta({\vec\rho}\,,\Omega)]
[1-i\delta({\vec\rho}\,,-\Omega)]-{\cal A}_p^2}\,,\label{U}
\\
V({\vec\rho}\,,\Omega)&=&\frac{2{\cal A}_p}
{[1+i\delta({\vec\rho}\,,\Omega)]
[1-i\delta({\vec\rho}\,,-\Omega)]-{\cal A}_p^2}\,.\label{V}
\end{eqnarray}
Here,
$\delta({\vec\rho}\,,\Omega)=\Delta-\Omega
+(\rho/\rho_0)^2$ is the local mismatch function, with
$\rho_0=f\sqrt{\lambda \gamma/\pi c}$
a characteristic transverse length.

Now, the quantities of interest can be calculated by using Eq.(\ref{eofa1}),
and some simplifying assumptions, i.e.:
$\lambda f/S_p^{1/2}$ much smaller than the typical 
scale of change of 
$U$, $V$ and $s$, so we can take the latter functions 
out of integral when 
they enter as a product with $\wp$;
the size of each pixel is the smallest of
all the spatial scales, so that we can substitute integration 
over the pixel area by multiplication
by $S_d$;
observation 
time $T_d$ long compared with the
inverse cavity bandwidth $\gamma^{-1}$. 
All that leads to
\begin{equation}\label{NI1}
\langle N_I({\vec\rho}\,,t)\rangle
=\eta S_d T_d s^2({\vec\rho}\,)
G({\vec\rho}\,)+{\rm noise}
\,,\end{equation}
with $G$ the gain factor
\begin{equation}\label{G1}
G({\vec\rho}\,)
=\{[(1+{\cal A}_p)^2-\delta({\vec\rho}\,,0)]^2+
4\delta^2({\vec\rho}\,,0)\}/\{[1+\delta^2({\vec\rho}\,,0)
-{\cal A}_p^2]^2\}
\,.\end{equation}
For the variance we get
\begin{eqnarray}\label{DeltaNI1}
\langle 
\Delta N_I^2({\vec\rho}\,,t)\rangle&=&\eta S_d T_d 
s^2({\vec\rho}\,)G({\vec\rho}\,)
\left\{1-\eta+\eta 
\left[ \cos^2\theta({\vec\rho}\,)e^{2R({\vec\rho}\,)}+\sin^2
\theta({\vec\rho}\,)e^{-2R({\vec\rho}\,)}\right]
\right\}\nonumber
\\&+&{\rm self\; interference\; of\; the\; noise}
\,,
\end{eqnarray}
where we have introduced the squeezing parameter 
$\exp\left[\pm R({\vec\rho}\,)\right]
=|U({\vec\rho}\,,0)|\pm |V({\vec\rho}\,,0)|$,
and the orientation 
angle 
$2\theta({\vec\rho}\,)=
\arg\left[U({\vec\rho}\,,0)+V({\vec\rho}\,,0)\right]
-\arg\left[U({\vec\rho}\,,0)\right]
-\arg\left[V({\vec\rho}\,,0)\right]$.
The condition to neglect the (unspecified) noise 
terms in Eqs.(\ref{NI1}), 
(\ref{DeltaNI1}) reads
\begin{equation}\label{noisecond}
s^2({\vec\rho}\,)(\lambda^2 f^2/S_p) 
(2\pi/\gamma)\gg 1\,.
\end{equation}
It also fixes the resolution of the scheme.
Once it is satisfied, the 
noise figure becomes
\begin{equation}\label{F1}
F=\{1-\eta+\eta[ 
\cos^2\theta({\vec\rho}\,)e^{2R({\vec\rho}\,)}+\sin^2
\theta({\vec\rho}\,)e^{-2R({\vec\rho}\,)}]\}/
\{\eta G({\vec\rho}\,)\}
\,.
\end{equation}
The optimum condition for noiseless amplification, coming from Eqs.(\ref{G1}) and (\ref{F1}),
is $\delta({\vec\rho}\,,0)=0$. Hence, depending on the value of the detuning, the noiseless
amplification  takes places in a small region around the optical axis or in a anular region.
Therefore, for reconstruction of the whole image one has to use scanning.

\section{Confocal cavity}

Let us now go to the configuration with confocal cavity \cite{our}. 
In this case the dashed part of the optical scheme in Fig.1 has to be considered.
In particular the confocality is guaranteed by means of a specific relation 
between the focal length of the intracavity lens and the cavity round trip length.
The intracavity field $b$ can be expanded on the basis 
of the Gauss-Laguerre modes $\{f_{p,l,i}({\vec\rho}\,)\}$ as
\begin{equation}\label{bGL}
b({\vec\rho}\,,t)=
{\sum_{p,l,i}}
f_{p,l,i}({\vec\rho}\,)\,b_{p,l,i}(t)\,.
\end{equation}
All the modes with $l$ even have the same frequency; the same 
is true for the modes with $l$ odd;
the frequency separation between the two groups of modes is 
equal to one half the free spectral
range.
Due to the frequency degeneracies, the field can be split 
into even and 
odd part
$b({\vec\rho}\,,t)
=b_+({\vec\rho}\,,t)
+b_-({\vec\rho}\,,t)$.
We consider the modes with $l$ even quasi-resonant with the signal.
Since the detuning is equal for 
all these modes, it is possible to get the following
Langevin equation for the intracavity field 
\begin{equation}\label{Langb2}
\partial_t\,b_+({\vec\rho}\,,t)=
-\gamma\left[(1+i\Delta_+)b_+({\vec\rho}\,,t)-
{\cal A}_p\,b_+^{\dag}({\vec\rho}\,,t)
\right]+\sqrt{2\gamma}\,b_+^{in}({\vec\rho}\,,t)\,.
\end{equation}
This equation can be 
solved in the frequency domain;
furthermore, with the aid of transformations at lenses
\begin{equation}\label{lenses2}
b_{in}({\vec\rho}\,,t)\equiv a({\vec\rho}\,,t)\,,
\quad
e({\vec\rho}\,,t)=
\int\, \frac{d{\vec\xi}}{\lambda f}\,
{\cal P}({\vec\xi}\,)\int\, \frac{d{\vec\rho}\,'}{\lambda f}\,
b_{out}({\vec\rho}\,'\,,t)\,
e^{i\frac{2\pi}{\lambda f}({\vec\rho}\,'-{\vec\rho}\,)
\cdot{\vec\xi}}\,,
\end{equation}
we can write
\begin{equation}\label{eofa2}
e({\vec\rho}\,,\Omega)=
\frac{1}{\lambda f}\int\, d{\vec\rho}\,'\,
\wp({\vec\rho}-{\vec\rho}\,')
\left[U(\Omega)a_+({\vec\rho}\,'\,,\Omega)
+V(\Omega)a_+^{\dag}({\vec\rho}\,'\,,-\Omega)
\right]\,,
\end{equation}
where the coefficients $U(\Omega)$ and $V(\Omega)$ are the same of Eqs.(\ref{U}) and (\ref{V})
with now the mismatch function no longer dependent from the position vector, i.e. 
$\delta(\Omega)=\Delta_+-\Omega$. 
Repeating the steps of previous Section, we 
easily obtain the mean number of
photoelectrons
\begin{equation}\label{NI2}
\langle N_I({\vec\rho}\,,t)\rangle=
\eta S_d T_d s^2({\vec\rho}\,)G
+{\rm noise}\,,
\end{equation}
where now the gain factor $G$ takes a simpler form
\begin{equation}\label{G2}
G=[(1+{\cal A}_p)/
(1-{\cal A}_p)]^2\,,
\end{equation}
considering the situation of perfect resonance, 
i.e. $\Delta_+=0$.
For the variance we have
\begin{equation}\label{DeltaNI2}
\langle \Delta N_I^2({\vec\rho}\,,t)\rangle=
\eta S_d T_d s^2({\vec\rho}\,)G\left\{
1-\eta
+\eta G \right\}
+{\rm self\; interference\; of\; the\; noise}
\,.
\end{equation}
To neglect the noise terms in Eqs.(\ref{NI2}) and (\ref{DeltaNI2}), we again 
use the condition (\ref{noisecond}). Then,
the noise figure becomes
\begin{equation}\label{F2}
F=\{1-\eta
+\eta G\}/
\{\eta G\}\,.
\end{equation}
As can be seen from Eqs.(\ref{G2}) and (\ref{F2}),  the noiseless amplification 
occours uniformely over the transverse plane.
Hence, this scheme offer the possibility of amplification of the whole image at once.
Moreover, a high gain is required to only
compensate the effect of non efficient detection.

\newpage

FIGURE CAPTIONS

Fig.1 A possible realization of the
parametric image amplifier.


\begin{thebibliography}{99}

\bibitem{noiseless}
C. M. Caves, Phys. Rev. D {\bf 26}, 1817 (1982);
B. Yurke and J. S. Denker, Phys. Rev. A {\bf 29}, 1419 (1984);
G. J. Milburn, {\it et al.}, {\it ibid.} {\bf 35}, 4443 (1987).

\bibitem{PLF94}
I. E. Protsenko, {\it et al.}, Phys. Rev. A {\bf 50}, 1627 (1994).

\bibitem{KL95}
M. I. Kolobov and L. A. Lugiato, Phys. Rev. A {\bf 52}, 4930 (1995).

\bibitem{our}
A. Gatti, L. A. Lugiato, S. Mancini and K. Petsas, to be submitted.





\end{thebibliography}
\end{document}